# Detecting software vulnerabilities using Language Models

Marwan Omar
College of Computing
Illinois Institute of Technology
Chicago, IL 30332–0250
Email: momar3@iit.edu

*Abstract*—Recently, deep learning techniques have garnered substantial attention for their ability to identify vulnerable code patterns accurately. However, current state-of-the-art deep learning models, such as Convolutional Neural Networks (CNN), and Long Short-Term Memories (LSTMs) require substantial computational resources. This results in a level of overhead that makes their implementation unfeasible for deployment in realtime settings. This study presents a novel transformer-based vulnerability detection framework, referred to as VulDetect, which is achieved through the fine-tuning of a pre-trained large language model, (GPT) on various benchmark datasets of vulnerable code. Our empirical findings indicate that our framework is capable of identifying vulnerable software code with an accuracy of up to 92.65%. Our proposed technique outperforms SyseVR and VulDeBERT, two state-of-the-art vulnerability detection techniques.
·

## I. Introduction

Cybersecurity seeks to safeguard machines from cyberattacks, which have become increasingly sophisticated and intense as businesses become more interconnected due to technological advancements. The ability of organizations and individuals to protect themselves from mass attacks is being questioned as the Verizon Cost of Data Breach Report 2023 shows that companies take an average of 197 days to detect a security breach and 69 days to contain it, risking significant financial and operational losses, unscheduled downtime, and lower productivity. It has become crucial for computers to process and analyze large amounts of language data, particularly in natural language interactions, and in cybersecurity applications such as flaw detection in software code. Traditional vulnerability detection methods, which rely on human experts to define vulnerabilities, are time-consuming and tedious. Machine learning techniques, particularly NLP models like CodeBERT, are capable of detecting software vulnerabilities without extensive feature engineering requirements, making detection faster and more automated.

Knowledge distillation (KD) is a technique used to reduce the size of neural networks so they can be run on devices with limited computing resources [1]. KD works by having a small, lightweight model (the "student") mimic the outputs of a larger network (the "teacher"), with the goal of transferring the teacher's high performance to the smaller architecture [2]. Interestingly, in some cases, the student model can even outperform the teacher, even when they have the same capacity [15,3,12]. This has been attributed to the presence of "dark knowledge," or hidden information about the teacher's learned representations that can be discovered through its outputs (such as class similarities) and utilized better by the student than the original labels [6].

In knowledge distillation, as depicted in Figure 1, a smaller "student" model is trained to imitate a larger "teacher" model and take advantage of the teacher's knowledge to achieve similar or better accuracy. In the following section, I will go into greater detail about the framework and components of knowledge distillation. The use of transformer-based models like GPT-2 for vulnerability detection offers several advantages, including improved accuracy and natural language processing capabilities. These models eliminate the need for manual input in static analysis tools and offer a faster, more automated detection process. The proposed vulnerability detection system called VulDetect harnesses the power, speed, and accuracy of Large Language Models (LLMs) based on transfers and utilizes GPT-2 models to detect vulnerabilities in C and C++ source code.

1) We propose VulDetect to detect software vulnerabilities leveraging the capabilities of large language models.
2) Utilizing benchmark datasets and a large language model based on GPT, We demonstrate that VulDetect can be used to identify vulnerable code in various programming languages including C/C++ and Java.
3) Our results show that VulDetect outperforms two current state-of-the-art techniques in detecting software vulnerabilities.

## II. RELATED WORK

Security researchers in academia and industry face significant challenges in detecting software vulnerabilities, despite the progress made using various source code analysis methods. Recent advancements in deep learning have prompted researchers to explore the effectiveness of these techniques in vulnerability detection. One such effort used CNNs and RNNs as feature extractors to detect vulnerabilities in raw source code. Another study introduced the concept of "code gadgets," which extracted library/API function calls from the source code. While graph-based approaches like Devign and DeepWukong utilized GNNs to learn data and control dependency code graphs, DeepTective aimed to detect SQL injection, XSS, and command injection vulnerabilities in PHP source code using a combination of GRUs and Graph Convolutional Networks. Additionally, researchers have addressed issues with imbalanced datasets through resampling techniques and curated benchmark datasets. Transformer-based models such as CodeBERT have been employed to automate the generation of program documentation and facilitate the search of natural language code. BERTBase, fine-tuned using a database of C/C++ source files, has also shown potential for detecting vulnerabilities. Finally, dynamic analysis techniques such as fuzzing and taint analysis have also been explored but face challenges with increasing runtime.

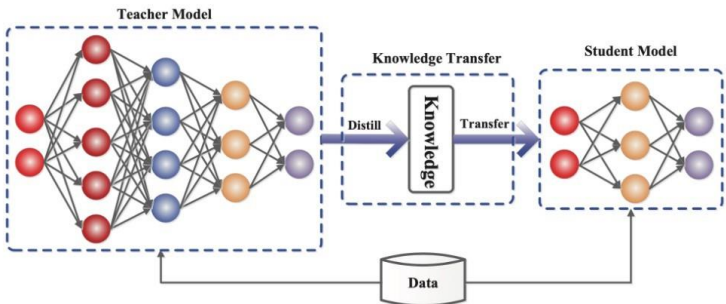

Fig. 1. An Overview of our Defense Framework

In a separate study, researchers investigated the effectiveness of transformer-based models for detecting software vulnerabilities. Fine-tuning the BERTBase model with a database of C/C++ source files, they achieved a detection accuracy of 93.49%, outperforming standard LSTM and BiLSTM models. The dataset used in this study was relatively small, and the authors used a limited number of models compared to the study at hand, which used multiple model architectures such as MegatronBERT and GPT-2. Furthermore, the researchers in this study leveraged code gadgets to input data, as opposed to removing labels and comments from the source file.

## III. METHODOLOGY

Algorithm for Online Knowledge Distillation: The input to the Algorithm for Online Knowledge Distillation for network that has been trained on a smaller dataset of labeled examples than the original teacher model, but has achieved a similar level of accuracy in detecting vulnerabilities.

Input: Set of labeled training data Output: A trained student model

Let $D$ be the dataset of software code, where each example is represented as a tuple $(x,y)$ where $x$ is a piece of software code and $y$ is a binary label indicating whether the code is vulnerable or not.

vulnerability detection is a set of labeled training data consisting of source code slices, where each slice represents a potential vulnerability. The input also includes a pre-trained teacher model, which is a larger and more accurate model used to generate soft labels for the training data. Finally, the input includes a smaller and less accurate student model, which is trained to predict the soft labels generated by the teacher model. The goal of the algorithm is to transfer knowledge from the teacher model to the student model through the use of distillation, thereby improving the student model's accuracy in detecting vulnerabilities. The output of the Algorithm for Online Knowledge Distillation for vulnerability detection is a trained student model that has learned to detect vulnerabilities in software code. Specifically, the student model is a neural Let $T$ be the set of $K$ teacher models, where each teacher model $T_k$ is trained on a different dataset or with a different configuration.

Let $S$ be the student model, which is a simpler and more lightweight model than the teacher models.

Let $\theta_T^{(k)}$ denote the parameters of the $k$-th teacher model, and $\theta_S$ denote the parameters of the student model.

The OKDDP algorithm can be outlined as follows:

Initialize the parameters of the student model $\theta_S$ randomly.

For each teacher model $T_k$, compute the predictions $p_k(x)$ for each example in the dataset $D$.

Initialize the student model's weights with the same values as the $k$-th teacher model.

Train the student model using a combination of the predictions from the teacher models as input, and the true labels $y$ as output. The loss function for the student model can be defined as a combination of the cross-entropy loss and the knowledge distillation loss: KDLoss$(\theta_S, \theta_T^{(k)}; D) = \frac{1}{n} \sum_{i=1}^{n} D_{\text{KL}}(p_k(x_i) || q_S(x_i; \theta_S, \theta_T^{(k)}))$

In this code, the Kullback-Leibler divergence is denoted by $D_{\text{KL}}$, and the softmax output of the student model is denoted by $q_S(x_i; \theta_S, \theta_T^{(k)})$. The notation for the parameters of the student and teacher models is the same as in the previous example.

## IV. DISTILLATION WITH A TEACHER MODEL

The technique of teacher-presented knowledge distillation involves training a student model using both soft and hard labels provided by the teacher. The soft labels, or targets, are calculated by applying the softmax function to the teacher's logits with a temperature parameter T. A higher T results in softer targets. In this study, we follow the suggestion of Lan, Zhu, and Gong (2018) and set T to 3 for all methods. To transfer knowledge from the teacher model to the student model, the predicted distribution q of the student model is aligned with the target distribution t, both calculated with a temperature of 3. The distillation loss function, which measures the difference between the two distributions, is represented by the Kullback-Leibler (KL) divergence equation (3). The total loss for training the model with both hard and soft labels is given by equation (4), where the distillation loss Ldis is multiplied by T2 before combination to maintain its contribution roughly unchanged. It's important to note that the probabilities predicted by the student model q are computed from the logits with a temperature of 1 when aligning with hard labels, and with a higher temperature when aligning with soft targets. Throughout the paper, the notation q represents the T=1 version and q represents the high-temperature version.

### A. Datasets

In this section, we present the datasets utilized in the remainder of this paper, which comprise function-level C/C++ source code obtained from diverse codebases such as opensource repositories and synthetic code samples. We categorize these datasets into two groups, based on their primary application for either pre-training or fine-tuning. It is important to note that all datasets mentioned in this section are publicly accessible and available for download.

*1) SARD:* The software assurance reference dataset (SARD) [8] is SARD is especially appealing because it includes security vulnerabilities as well as nonvulnerable alternatives, which allows our models to learn the differences between security vulnerabilities. Afterward, we perform a preprocessing step in order to remove any unwanted artifacts that could lead to the overfitting of our models.

*2) SeVC:* Semantics-based Vulnerability Candidate [5]: This dataset Contains 1,591 C/C++ open-source programs from the NVD, as well as 14,000 open-source programs from the SARD. There are 420,627 SeVCs in it, with 56,395 of them being vulnerable, and 364,232 of them not being vulnerable. Moreover, there are four types of SyVCs: Library/API Function Calls, Array Usage, Pointer Usage, and Arithmetic Expression.

### B. Devign

The Devign dataset, which was first introduced in [9], is a real-world dataset for detecting software vulnerabilities. The dataset includes function-level C/C++ source code from two widely-used open-source software projects, QEMU and FFmpeg. The labeling and verification process was carried out by a team of security researchers in two rounds, and was done manually.

*1) D2A:* The IBM Research team introduced and curated the D2A dataset for real-world vulnerability detection [7]. This dataset includes various open-source software projects such as FFmpeg, httpd, Libav, LibTIFF, Nginx, and OpenSSL. To label the issues reported by static analyzers, the team utilized a differential analysis technique.

## V. EVALUATION AND RESULTS

%beginitemize

### A. Configuration

We Implemented our experiments using an ASUS TUF Gaming laptop with an Intel Core i7-8th generation CPU. The processor has six cores with a maximum operating frequency of 2.2 GHz for each core. sectionResult

| Model Score | SARD | SeVC | Devign | D2A | |
|---|---|---|---|---|---|
| VulBERTa | 88.7 | 84.2 | 80.5 | 81.8 | 79.9 |
| SyseVR | 81.5 | 82.6 | . | 78.3 | 80.2 72.7 |
| DistilVulBERT | 94.0 | 91.4 | 82.2 | 87.5 | 85.9 |

### B. Defense Performance Evaluation

To evaluate the performance of our technique, we decided to test it on the same three classifiers (GPT-2, CodeBERT, and LSTM) using the same two benchmark datasets (SARD, and SeVC). The goal was to see how effective our technique is in detecting Vulnerabilities. Table I illustrates that models' classification accuracy is significantly higher when the VulDetect technique is implemented. In particular, we observe that our GPT-2 classifier fared the best with a classification accuracy of up to 92.59 when tested on the SARD dataset. On the other side of the spectrum, we noticed that LSTM fared the worst with a classification accuracy of 65.78 when evaluated on theSeVC dataset. To sum up, we observe that GPT-2 is consistently performing the best among the other two network architectures (CodeBERT, and LSTM). This is not surprising as GPT is a transformer-based learning model and is known to achieve state-of-the-art performance under various linguistic tasks including sentiment analysis and sentence classification. Table 2: DistilBERT yields comparable performance on downstream tasks. Comparison on downstream tasks: IMDb (test accuracy) and SQuAD 1.1 (EM/F1 on dev set). D: with a second step of distillation during fine-tuning.

## VI. Experiments

We assess the performance and generalization capabilities of DistilVulBERT on 4 popular benchmark datasets for detecting software vulnerabilities. We report scores on the development sets for each dataset by fine-tuning DistilVulBERT without the use of ensembling or a multi-tasking scheme for fine-tuning. We compare the results to the baseline provided by the authors
of VulBERTa [3] [4]

The results on each of the four datasets are shown on Table 1 along with the macro-score (average of individual scores). Among the 9 tasks, DistilBERT is always on par or improving over the ELMo baseline (up to 19 points of accuracy on STS-B). DistilBERT also compares surprisingly well to BERT, retaining 97% of the performance with 40% fewer parameters.

### A. Baseline Comparison

We evaluate the performance of VulDetect at detecting vulnerabilities using the three models (GPT-2, BERT, and LSTM) and observe that VulDetect outperforms VulDeBERT in all cases. In particular, we notice that our technique achieves up to 92.4 $F1$ score on the SARD dataset for the GPT2 model. This shows that our VulDetect solution is most effective in detecting vulnerabilities. On the other side of the spectrum, we observe that our technique fared the worst on the LSTM model. In particular, we show that VulDetect technique detected only a portion of the vulnerabilities, with 53.5 $F1$ score. We believe that our LSTM classifier may have suffered from overfitting on the dataset, a drawback of LSTM models in-general.

## VII. Conclusion

The utilization of language models that possess high power, speed, and accuracy is essential for the progression of the domain of cybersecurity. This is particularly relevant in regard to proactively identifying software vulnerabilities to enhance security solutions. In the current study, we introduce VulDetect, a cutting-edge vulnerability detection framework that leverages the capabilities of deep learning and language models. To evaluate the efficacy of our proposed method, we conducted experiments in which we trained several state-of-the-art NLP architectures on two benchmark datasets (SARD and SeCV) for vulnerability identification. The results demonstrate that the proposed VulDetect outperforms the SySeVR and VulDeBERT techniques in detecting software vulnerabilities.

[10]